\documentclass[prb,aps,reprint,showpacs,preprintnumbers,citeautoscript,amsmath,amssymb,showkeys,superscriptaddress]{revtex4-1}

\usepackage{graphicx}
\usepackage{multirow}
\usepackage{graphicx}
\usepackage{dcolumn}
\usepackage{bm}
\usepackage{lipsum}
\usepackage[most]{tcolorbox}
\usepackage[section]{placeins}
\usepackage{float}
\usepackage{mwe}
\usepackage{amsmath,amssymb}
\usepackage{amsmath}

\begin{document}

\title{Weyl semimetal phases and intrinsic spin-Hall conductivity in SbAs ordered alloys}

\author{Muhammad Zubair} 
\email{mzubair@udel.edu}
\affiliation{Department of Physics and Astronomy, University of Delaware, Newark, DE 19716, USA}
\affiliation{Department of Materials Science and Engineering, University of Delaware, Newark, DE 19716, USA}

\author{Dai Q. Ho}
\affiliation{Department of Materials Science and Engineering, University of Delaware, Newark, DE 19716, USA}
\affiliation{Faculty of Natural Sciences, Quy Nhon University, Quy Nhon 55113, Vietnam}

\author{Duy Quang To}
\affiliation{Department of Materials Science and Engineering, University of Delaware, Newark, DE 19716, USA}

\author{Shoaib Khalid}
\affiliation{Princeton Plasma Physics Laboratory, P.O. Box 451, Princeton, New Jersey 08543, USA}

\author{Anderson Janotti}
\email{janotti@udel.edu}
\affiliation{Department of Materials Science and Engineering, University of Delaware, Newark, DE 19716, USA}

\begin{abstract}
Using density functional theory calculations we investigated possible Weyl semimetal (WSM) phases in antimony arsenide ordered alloys Sb$_{1-x}$As$_x$ ($x=0, 1/6, 1/3, 1/2, 2/3, 5/6, 1$). We find WSM phases for all As compositions of Sb$_{1-x}$As$_x$ with broken inversion symmetry, in contrast to Bi$_{1-x}$Sb$_x$ where only compositions $x=1/2$ and $5/6$ were predicted to exhibit WSM phases. The WSM phases in Sb$_{1-x}$As$_x$ are characterized by the presence of 12 Weyl points, located within 55 meV from the Fermi level in the case of $x$=1/2. The robust spin-orbit coupling strength and Berry curvature in these alloys produce large spin-Hall conductivity in the range of 176-602 ($\hbar/e$)(S/cm), comparable to that in the BiSb alloys. Finally, Sb$_{0.5}$As$_{0.5}$ is predicted to be almost lattice-matched to GaAs(111), with the Fermi level within the gap of the semiconductor, facilitating growth and characterization, and thus, offering promising integration with conventional semiconductors. 
\end{abstract}


\maketitle
\section{Introduction} \label{sec:intro}

Weyl semimetals (WSMs) have attracted significant interest in condensed matter physics due to their unique electronic and magnetic properties, characterized by linear band dispersions near Weyl points where conduction and valence bands intersect \cite{burkov2011weyl,wan2011topological,balents2011weyl,xu2015observation}. These materials exhibit exotic phenomena such as chiral anomaly observed via negative magnetoresistance, unusual surface states (Fermi arcs), and high charge carrier mobility \cite{Burkov2018,huang2015observation,zhang2016signatures,wang2016helicity,wang2016helicity,shekhar2015extremely,ghimire2015magnetotransport,moll2016magnetic,xu2011chern,sun2016strong,liu2018giant,gooth2017experimental,burkov2015chiral,hasan2017discovery}, making them promising candidates for future quantum electronic and spintronic devices.  WSM phase is expected to exist in materials with broken time-reversal symmetry, broken inversion-symmetry, or both. Many WSMs have been predicted by density functional theory calculations \cite{su2018topological,su2018topological,singh2016prediction,luo2015electron,shi2018prediction} and later verified experimentally through angle-resolved photoemission spectroscopy (ARPES) and scanning tunneling microscopy (STM). Most of the studies were carried out in materials lacking inversion center \cite{weng2015weyl,meng2019dirac,yang2015weyl}.

Alloying allows for breaking inversion symmetry, opening an avenue for the realization of Weyl semimetal phases in topological materials. For instance, Bi$_{1-x}$Sb$_{x}$ was one of the first experimentally identified 3D topological semiconducting alloys using ARPES \cite{hsieh_topological_2008,nishide2010,hirahara2010}.  A transition from topologically trivial to non-trivial phase occurs at the concentration of $x=0.04$, turning Bi$_{1-x}$Sb$_{x}$ alloys into topological semimetal with an inverted band gap. Violation of Ohm's law and chiral anomaly-induced negative magnetoresistance were observed recently in the transport measurements as signatures of WSM phase in Bi$_{1-x}$Sb$_{x}$ \cite{shin2017violation,kim2013dirac,singh2018elastic}.
Su {\rm et al.}  \cite{su2018topological}investigated Weyl semimetal phases in Bi$_{1-x}$Sb$_{x}$ ($x=$0.17, 0.33, 0.5, 0.67, 0.83) ordered alloys exploring combinations of atomic composition and arrangement using first-principles calculations\cite{su2018topological}. A Weyl semimetal phase was found for Sb concentrations $x=0.5$ and $x=0.83$, indicating that chemical composition and specific atom arrangements are key to the existence of the topological phase. 

\begin{figure}
\centering
\includegraphics[width=3in]{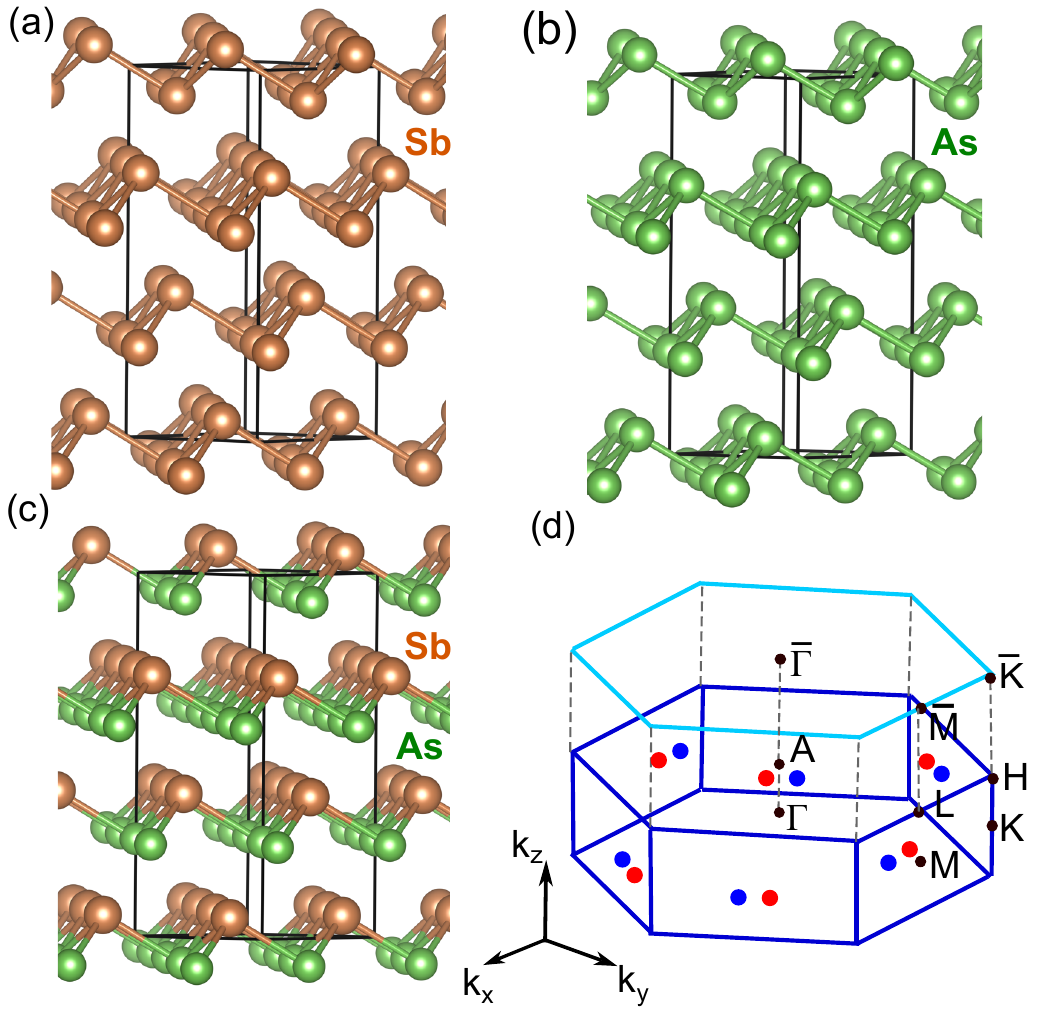}
\caption{Layered hexagonal crystal structure of (a) Sb and (b) As using the conventional unit cell with 6 atoms, R$\bar{3}$m space group,
(c) Sb$_{0.5}$As$_{0.5}$ ordered alloy with R3m space group, and (d) correspondingly the 3D Brillouin zone of the hexagonal cell and the projected 2D Brillouin zone of the (0001) surface Brillouin zone showing the high symmetry points and directions. The location of the Weyl points with positive chirality (red) and negative chirality (blue) are also illustrated. }
\label{fig1}
\end{figure}

\begin{figure*}
\centering
\includegraphics[width=5in]{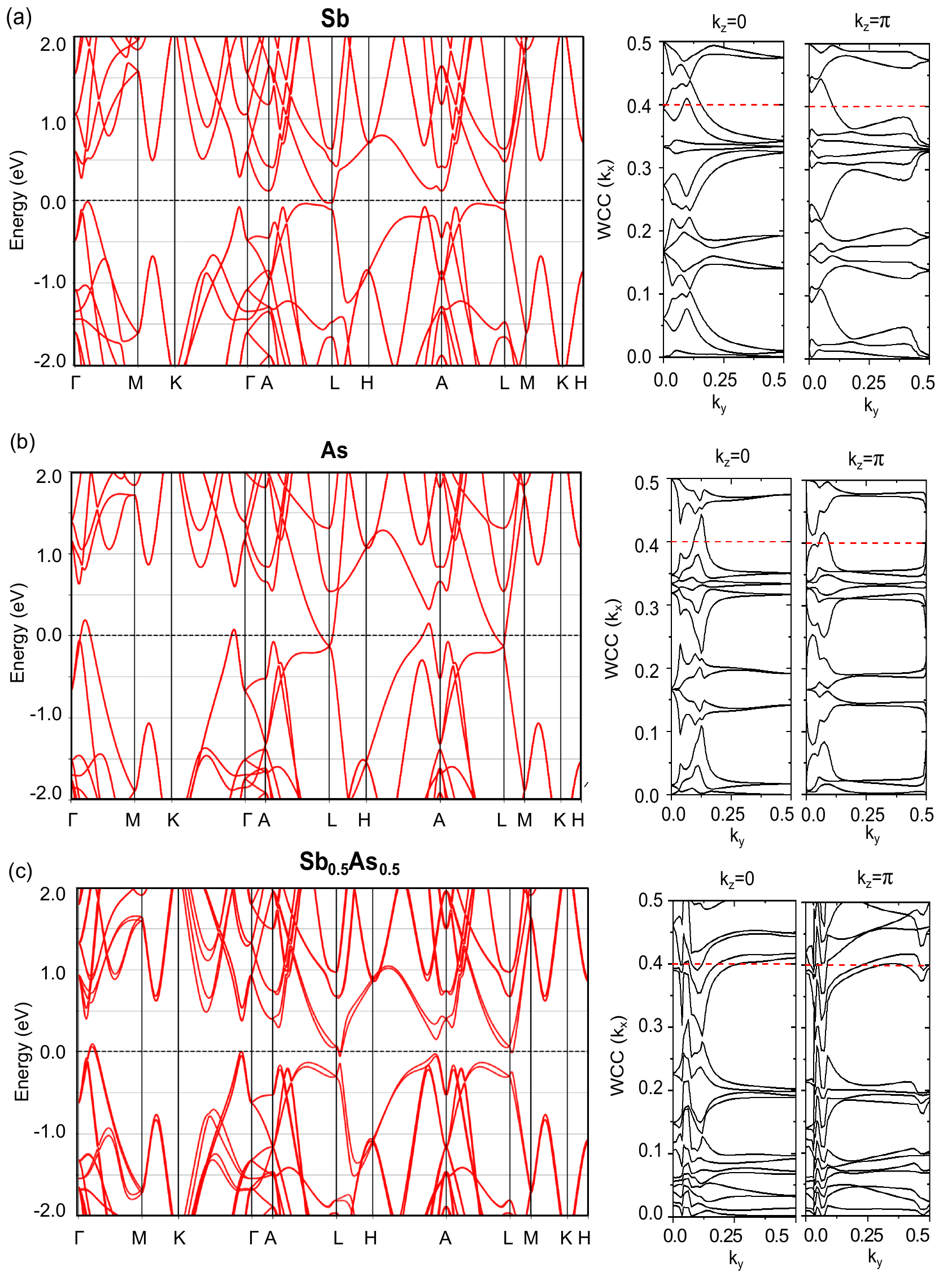}
\caption{Electronic band structure and Wannier charge centers (WCC)/Wilson loop on k$_z$=0 and k$_z$=$\pi$ planes for (a) Sb, (b) As and, (c) Sb$_{0.5}$As$_{0.5}$ ordered alloy. The evolution of the Wannier charge center/Wilson loop crosses the red dashed line an even number of times on K$_z$=0 plane and an odd number of times on K$_z$=$\pi$ plane in (a), whereas in (b) and it crosses the red dashed line an even number of times on k$_z$=0 and K$_z$=$\pi$ plane. The red dashed line in (c) crosses the Wannier charge center/Wilson loop an even number of times on k$_z$=0 plane and an odd number of times on k$_z$=$\pi$ plane.}
\label{fig2}
\end{figure*}

Inspired by the successful identification and characterization of topological properties in BiSb alloys, here we investigate SbAs ordered alloys as possible Weyl semimetals. Similar to BiSb, Sb$_{1-x}$As$_x$ alloys offer an adjustable electronic structure that can potentially host Weyl points under specific conditions. Furthermore, SbAs alloys, with their closely related electronic properties to BiSb, present a new avenue for exploring Weyl semimetal behavior in, perhaps, a more versatile composition range. Despite replacing Bi with As could lead to weaker spin-orbit coupling, stronger bonding and shorter bond lengths in SbAs alloys are possibly translated into higher stability with respect to BiSb and, thus, facilitate synthesis and characterization. Furthermore, the integration of these alloys with conventional III-V semiconductors, like GaAs, InAs, and InSb, opens up exciting possibilities for novel device architectures. III-V semiconductors are the backbone of modern optoelectronics, high-speed electronics, and photonics. Therefore, combining the topologically protected states of Weyl semimetals with the well-established properties of III-V compounds could lead to the development of hybrid devices with enhanced functionalities, such as topological transistors and low-dissipation electronics.

This paper aims to explore the potential of SbAs alloys as Weyl semimetals and their integration prospects with conventional III-V semiconductor materials by focusing on their lattice match and band alignments. By leveraging insights gained from BiSb alloys, we investigate the conditions (chemical composition and ordering) for SbAs to display WSM phases, seeking to advance the understanding of topological semimetals and their potential technological applications and integration into semiconductor-based platforms.

\section{Computational Approach} \label{sec:Com}

Density functional theory (DFT) \cite{hohenberg1964inhomogeneous,kohn1965self} calculations as implemented in the VASP code \cite{kresse1996efficient} were employed to investigate the structural and electronic properties of selected ordered structures representing Sb$_{1-x}$As$_x$ alloys. The interactions between the valence electrons and ions were treated using the projector augmented-wave (PAW) method \cite{kresse1999ultrasoft}, which include five valence electrons for Sb ($5s^25p^3$) and As ($4s^24p^3$). We used the generalized gradient approximation (GGA) including van der Waals interactions \cite{PhysRevLett.77.3865} for determining equilibrium structures and the electronic structures. Plane-wave basis set with a cutoff of 300 eV and a 13$\times$13$\times$5 $\Gamma$-centered k-mesh for the integration over the Brillouin zone of the 6-atom hexagonal cell were used in all calculations. For convergence of the electronic self-consistent calculations, the total energy difference criterion was set to $10^{-8}$ eV, and atomic positions were relaxed until the Hellmann-Feynman forces were less than $10^{-4}$ eV/\AA. The effects of spin-orbit coupling (SOC) were included in all electronic band structure calculations. To analyze the topological properties of Sb$_{1-x}$As$_x$ alloys, we projected the Bl\"och wave functions into maximally localized Wannier functions (MLWFs). The tight-binding Hamiltonian parameters are determined from MLWFs overlap matrix obtained using the wannier90 code\cite{mostofi2008wannier90}. 
Wannier Tools is used to determine the Berry curvature, Fermi arc, Weyl points, and the chirality of the Weyl points\cite{WU2017}. The intrinsic spin Hall conductivity is calculated using the wannier90 code\cite{mostofi2008wannier90}, and the mcsqs code from the Alloy Theoretic Automated Toolkit (ATAT) was used to generate special quasirandom structures (SQS) 
for simulating the Sb$_{0.5}$As$_{0.5}$ random alloys containing 48 atoms supercell. This method uses an annealing loop with an objective function simulated in Monte Carlo that looks for perfectly matched correlation functions in order to generate periodic supercells that mimic real disordered structures\cite{zunger1990special,van2013efficient}.

\section{Results and Discussion}\label{sec:result}


Both antimony (Sb) and arsenic (As) crystallize in the rhombohedral A7 crystal structure with space group R$\overline{3}$m, composed of two interpenetrating and diagonally distorted face-centered cubic (FCC) lattices. Sb (As) atoms are stacked along the (111) direction containing two atoms per primitive cell. It can also be described as a hexagonal unit cell with 6 atoms forming three bilayers that are weakly bonded by van der Waals interaction as shown in Fig.~\ref{fig1}(a,b). The Brillouin zone of the hexagonal unit and high symmetry points are shown in Fig.~\ref{fig1}(d). The calculated lattice parameters for Sb (hexagonal unit cell) are $a=4.362${\AA} and $c=11.210${\AA}, compared to the experimental values 4.299 {\AA} and  11.251 {\AA} \cite{Jette1935}, and for As they are $a=3.810${\AA} and $c=10.356${\AA}, compared to experimental values 3.760 {\AA} and 10.441 {\AA} \cite{Schiferl1969}.
Sb and As exhibit similar electronic band structures as shown in Fig.~\ref{fig2}(a,b). However, due to the different strengths in spin-orbit coupling (SOC), Sb and As have different band ordering and, therefore, different Z\textsubscript{2} topological invariants as shown on the right in Fig.~\ref{fig2}(a,b). We find Z\textsubscript{2}=1 for Sb and Z\textsubscript{2}=0 for As, characterizing Sb as a non-trivial topological semimetal and As a trivial semimetal. 

For a direct comparison of the electronic band structure of the Sb$_{1-x}$As$_x$ alloys with that of bulk Sb and As, we limited the scope of the present study by considering only ordered structures that can be described using the six-atoms hexagonal unit cell, varying the concentration in the range $x$= 0, 1/6, 1/3, 1/2, 2/3, 5/6 and 1. For each concentration, we consider all possible atomic arrangements, shown in Fig.~S1 in the Supplemental Material \cite{Zubair_supplement}. 

The calculated mixing enthalpy $\Delta H_f$, defined as the difference between the total energy of the alloy and the weighted sum of the total energy of the elemental phases, shows that only the structure with $x=0.5$ shown in Fig.~\ref{fig1}(c) has a negative value ($\Delta H_f=-9$ meV/atom), indicating stability at T=0. This structure consists of an alternating arrangement of Sb and As monatomic layers along the $c$ axis. The two other possible structures for $x=0.5$, shown in  Fig.~S1(c)-(II,III) have $\Delta H_f=31$ meV/atom and $\Delta H_f=40$ meV/atom, respectively. The structure in  Fig.~S1(c)-II consists of two monatomic Sb and As layers followed by an alternating monatomic Sb and As layer along the $c$ direction, and  Fig.~S1(c)-III consists of an alternating arrangement of three monatomic Sb and three As layers along $c$.
Using a supercell containing 48 atoms (2x2x2 repetition of the hexagonal 6-atoms cell), we used a special quasi-random structure \cite{sqs} to simulate a random alloy (Fig.S2 in the Supplemental Material \cite{Zubair_supplement}, and found that the ordered structure in Fig.~\ref{fig1}(c) is lower in energy than the random structure by 38 meV/atom.  Thus, the calculated $\Delta H_f$ for all composition $x$ indicate that Sb$_{1-x}$As$_x$ alloys energetically prefer to order at $x$=0.5, and that Sb and As prefer to sit in alternating planes of atoms in each bilayer.

\begin{figure}
\centering
\includegraphics[width=3.4in]{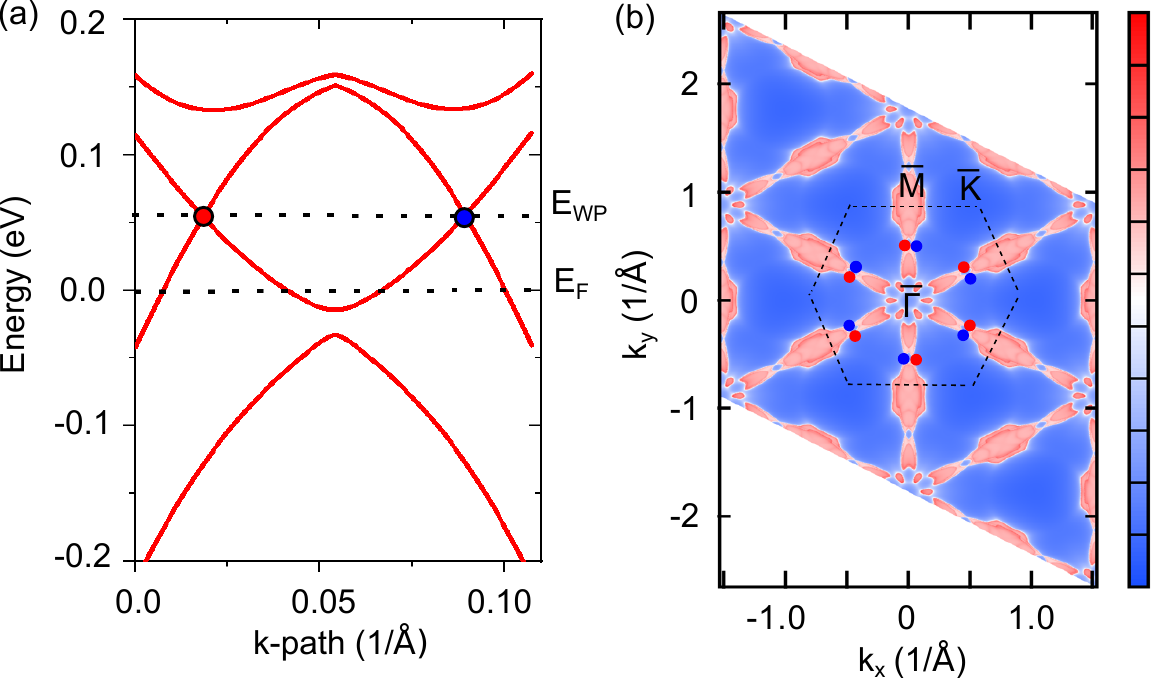}
\caption{Electronic bands passing through two Weyl points for (a) Sb$_{0.5}$As$_{0.5}$ ordered alloy, where E\textsubscript{F} represents the Fermi energy while E\textsubscript{WP} represents the energy of the Weyl points. (b) Fermi arc on the (0001) plane of Sb$_{0.5}$As$_{0.5}$ ordered alloy at the energy of the Weyl points E\textsubscript{WP}. The red circles indicate the source of Berry curvature and blue circles indicate the sink of Berry curvature.}
\label{fig3}
\end{figure}

We note that all the ordered Sb$_{1-x}$As$_x$ alloys where inversion symmetry is broken are predicted to be Weyl semimetals.  However, in the following we limit our discussion to the electronic structure of the lowest energy structure shown in Fig.~\ref{fig1}(c). Its electronic band structure is shown in Fig.~\ref{fig2}(c). The spin degeneracy is lifted for all bands due to the absence of inversion symmetry. To determine the topological phase, the Wannier charge center/Wilson loop is calculated in the high-symmetry planes k\textsubscript{z}=0 and k\textsubscript{z}=$\pi$ and displayed on the right in Fig.~\ref{fig2}(c). The Z\textsubscript{2} topological invariant $\bm{\nu}\textsubscript{0}$ for k\textsubscript{z}=0 and k\textsubscript{z}=$\pi$ plane is 0 and 1, demonstrating a topological nontrivial phase of the Sb$_{0.5}$As$_{0.5}$ ordered alloy.

\begin{figure}
\centering
\includegraphics[width=3.4in]{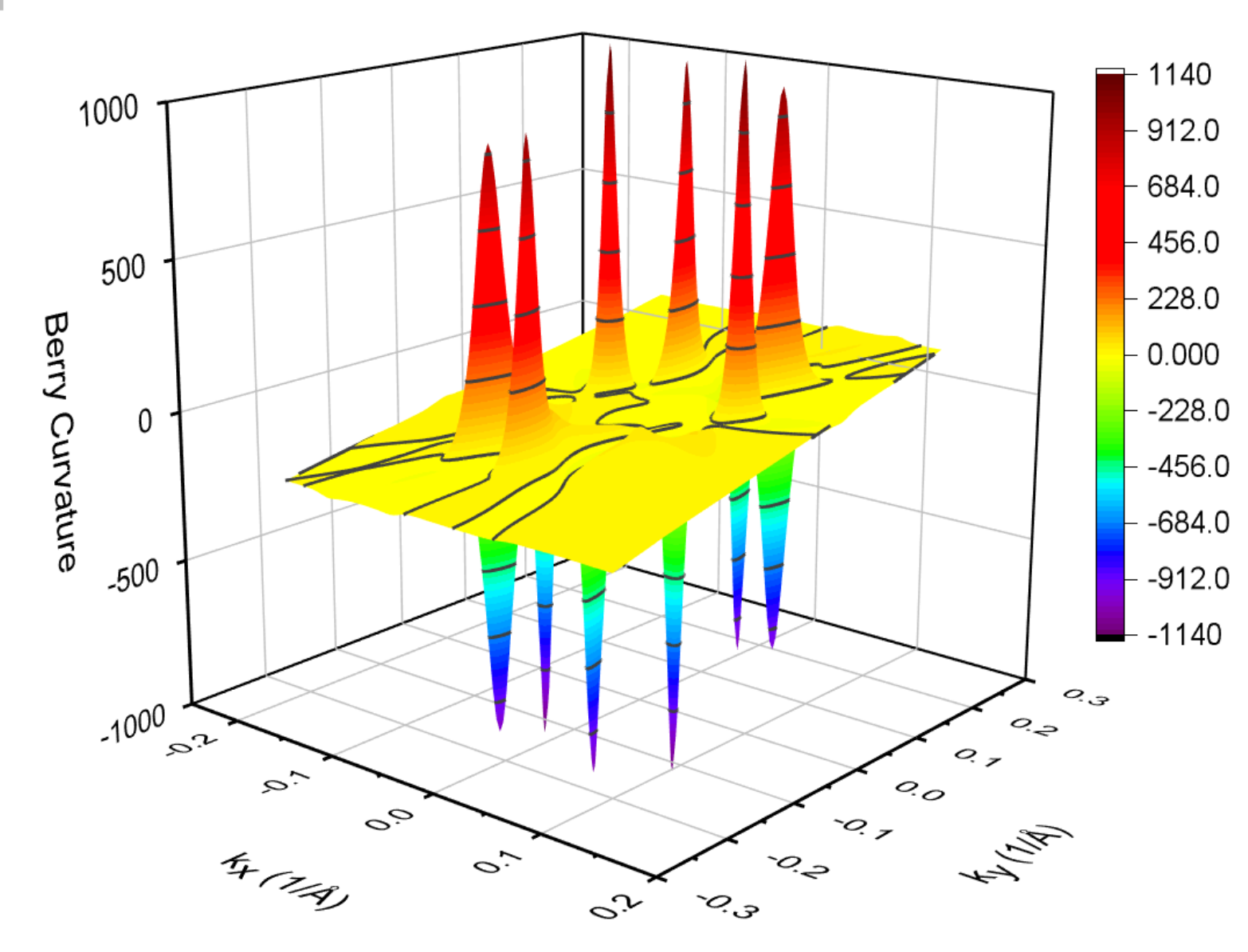}
\caption{Berry curvature distribution for Sb$_{0.5}$As$_{0.5}$ ordered alloys in the k$_{z}$=0 plane.}
\label{fig4}
\end{figure}

\begin{table}
\setlength{\tabcolsep}{12pt}
\caption{Position (in 1/{\AA}) of the 12 Weyl points (WPs) in the first Brillouin zone of the hexagonal unit cell of the lowest energy Sb$_{0.5}$As$_{0.5}$ ordered alloy (structure in Fig.~\ref{fig2}(c)). The chirality of each WP is also indicated.}
\centering
\begin{tabular}{r c c c}
\hline
k\textsubscript{1}  &   k\textsubscript{2}     &   k\textsubscript{3} & Chirality\\  
\hline 
$\phantom{-}0.019$ & $\phantom{-}0.288$ &  $\phantom{-}0.400$ & $-1$  \\
$\phantom{-}0.288$ & $\phantom{-}0.019$ &  $-0.400$& $\phantom{-}1$  \\
$-0.019$  & $\phantom{-}0.308$  &  $\phantom{-}0.400$  &  $\phantom{-}1$  \\
$\phantom{-}0.308$  &  $-0.019$  &  $-0.400$  &  $-1$  \\
$\phantom{-}0.019$  &  $-0.308$  &  $-0.400$  &  $\phantom{-}1$  \\
$-0.019$  &  $-0.288$  &  $-0.400$  &  $-1$  \\
$-0.308$  &  $\phantom{-}0.019$  &  $\phantom{-}0.400$  &  $-1$  \\
$-0.288$ &  $\phantom{-}0.308$ & $-0.400$  &  $-1$  \\
$\phantom{-}0.288$ & $-0.308$ &  $\phantom{-}0.400$  &  $-1$  \\
$-0.288$ & $-0.019$ &  $\phantom{-}0.400$  &   $\phantom{-}1$  \\
$\phantom{-}0.308$ & $-0.289$ &  $\phantom{-}0.400$  &  $\phantom{-}1$  \\
$-0.308$ &  $\phantom{-}0.289$ & $-0.400$  &  $\phantom{-}1$  \\[0.5ex]
\hline\hline 
\end{tabular}
\label{WPcoordinates} 
\end{table}

Although Sb$_{0.5}$As$_{0.5}$ is classified as semimetal, the band structure does not exhibit any band crossing along the high symmetry lines. We observed a pair of linear band crossing away from high symmetry points at (0.01986,0.28899,0.40084)(1/\AA) and (-0.01989,0.30884,0.40072)(1/\AA) as shown in Fig.~\ref{fig3}(a). The linear band crossing points in this pair are Weyl cones behaving as source and sink of Berry curvature. Since this structure exhibits C\textsubscript{3} rotation and time-reversal symmetry, there are a total of six pairs of Weyl points located in the entire Brillouin zone. The location and chirality of these Weyl points are listed in Table I. The Weyl points are located only 55 meV above Fermi level, facilitating their detection through surface sensitive probes such as ARPES, or bulk transport measurements. The Weyl points exhibit a linear touching in the projected bulk states, and the Fermi arcs are curved, which is a signature of type-II Weyl semimetal \cite{soluyanov2015type}.

\begin{table}
\caption{Comparison of the Spin Hall conductivities for different Materials.}
\centering
\begin{tabular}{l c c }
\hline\hline
Material  &   SHC($\hbar$/e) (S/cm)      \\  
\hline 
SbAs alloys  &  176-602& present work\\
Sb$_{0.5}$As$_{0.5}$ & 330 & present work \\
Bi & 772-1034 &  \cite{qu2023intrinsic}(Cal.)\\
Bi$_{0.85}$Sb$_{0.15}$ &  520 &  \cite{sharma2021light}(Cal.)\\
Bi$_{2}$Se$_{3}$ &  54-324 &  \cite{farzaneh2020intrinsic}(Cal.)\\
Bi$_{2}$Te$_{3}$ &  436-812 &  \cite{farzaneh2020intrinsic}(Cal.)\\
Sb$_{2}$Se$_{3}$ &  188 -226&  \cite{farzaneh2020intrinsic}(Cal.)\\
Sb$_{2}$Te$_{3}$ &  200-308 &  \cite{farzaneh2020intrinsic}(Cal.)\\
Al & 10-36 &  \cite{valenzuela2006direct,valenzuela2007electrical}(Exp.)\\
Silicon & 0.02&  \cite{ando2012observation}(Exp.)\\ 
GaAs & 0.02&  \cite{matsuzaka2009electron,tse2006spin}(Exp.)\\ 
GaAs & 0.0009&  \cite{engel2005theory}(Cal.)\\
ZnSe & 0.01&  \cite{stern2006current}(Exp.)\\
\hline\hline 
\end{tabular}
\label{tab1} 
\end{table}

The presence of Fermi arc is an important characteristic of Weyl semimetals \cite{balents2011weyl,wan2011topological,yan2017topological,potter2014quantum, xu2015observation}. 
The Fermi arc on the (0001) plane of the Sb$_{0.5}$As$_{0.5}$ ordered alloy at Weyl points energy is shown in Fig.~\ref{fig3}(b), located at 55 meV above the Fermi energy. The positive and negative chirality of the pairs of Weyl points are indicated and correspond to the source (red) and sink (blue) of Berry curvature. As evident from the Fermi arc plot, there are a total of 12 points located in the first Brillouin zone of the hexagonal unit cell. 


\begin{figure}
\centering
\includegraphics[width=3in]{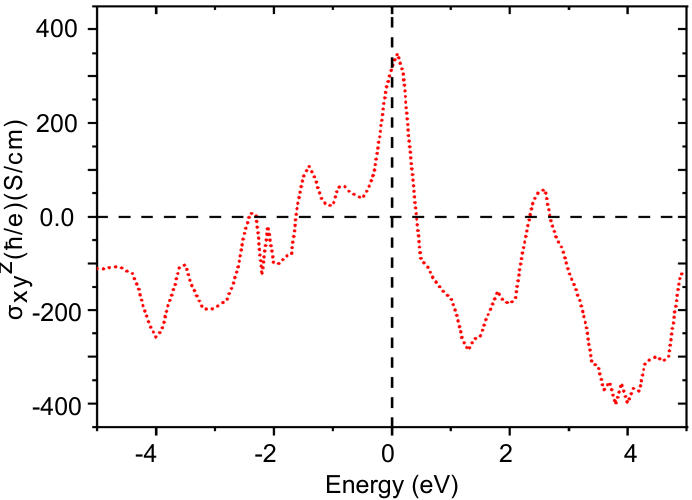}
\caption{Intrinsic spin-Hall conductivity in Sb$_{0.5}$As$_{0.5}$ ordered alloy.}
\label{fig5}
\end{figure}

\begin{figure}
\centering
\includegraphics[width=3in]{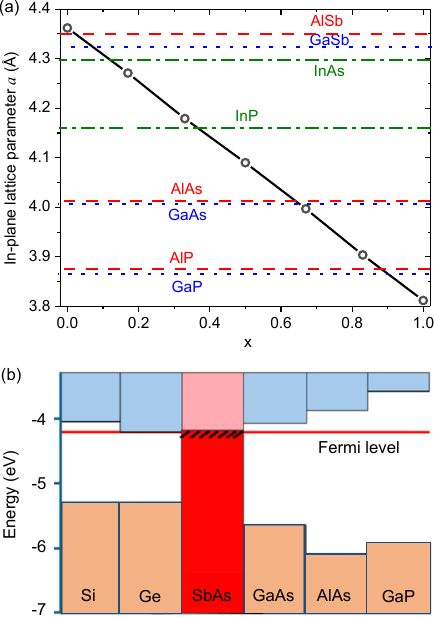}
\caption{(a) In-plane lattice parameter $a$ of Sb$_{1-x}$As$_x$ alloys as a function of composition $x$. In-plane (111) lattice parameters of III-V compounds are highlighted ($a_{(111)}=a/\sqrt(2)$, where $a$ is the cubic lattice parameter of the III-V).
(b) Calculated Fermi level of Sb$_{0.5}$As$_{0.5}$ ordered alloy with respect to the vacuum level and compared with that of conventional semiconductors from Ref.~\onlinecite{hinuma2014band}.
}
\label{fig6}
\end{figure}

The Berry curvature distribution is calculated using the Kubo formula \cite{xiao2010berry} given by
\begin{equation}
\label{berry_cur}
\Omega_{n}^{\gamma} = 2i\hbar^2 \sum_{m \neq n} \frac{\langle u_n(\vec{k})|\hat{v_{\alpha}}|u_m(\vec{k})\rangle \langle u_m(\vec{k})|\hat{v_{\beta}}|u_n(\vec{k})\rangle}{(E_n(\vec{k}) - E_m(\vec{k}))^2},
\end{equation} 
where $\Omega_{n}^{\gamma}$ is the Berry curvature for band $n$, $\hat{v}_{\alpha} = \frac{1}{\hbar}\frac{\partial \hat{H}}{\partial k_{\alpha (\beta, \gamma)}}$ is the velocity operator for $\alpha, \beta, \gamma = x, y, z$ and $|u_n(\vec{k}) \rangle$ and $E_n(\vec{k})$ are the eigenvectors and eigenvalues of the Hamiltonian $\hat{H}(\vec{k})$. We calculated the Berry curvature by taking the summation in $\Omega_{n}^{\gamma}$ over all the bands up to the lower band crossing in Fig.~\ref{fig3}(a). The Berry curvature distribution for $k_y$=0 plane is shown in Fig.~\ref{fig4}, where we can see various discontinuities. 
The positive spikes are related to the source of the Berry curvature, whereas the negative spikes are related to the sink of the Berry curvature. Six pairs of Weyl points are clearly identified, and their coordinates are listed in Table~\ref{WPcoordinates}.

\begin{table}
\setlength{\tabcolsep}{12pt}
\caption{Energies of Weyl points along with their composition in   SbAs alloys.}
\centering
\begin{tabular}{l c}
\hline\hline
Compositions  &   Energy of the Weyl points E\textsubscript{WP}(eV)\\  
\hline 
Sb$_{0.83}$As$_{0.17}$  &  -0.190\\
Sb$_{0.67}$As$_{0.33}$ & -0.090\\
Sb$_{0.5}$As$_{0.5}$&  0.055\\
Sb$_{0.33}$As$_{0.67}$ &  -0.020\\
Sb$_{0.83}$As$_{0.17}$  &  -0.116\\[0.5ex]
\hline\hline 
\end{tabular}
\label{lattice} 
\end{table}

We also calculated the intrinsic DC spin-Hall conductivity (SHC) using the Kubo-Greenwood formula\cite{qiao2018calculation}, given by:
\begin{equation} 
\begin{aligned}
 &\sigma_{xy}^{spinz}(\omega) = \hbar\int_{BZ}{\frac{d^3k}{(2\pi)^3}} \sum_{n}f\textsubscript{nk} \\
& \times \sum_{m \neq n} \frac {2Im[\langle nk|\hat{j}_x^{spinz}|mk\rangle \langle mk|-e\hat{v_{y}}|nk\rangle}{(\epsilon_{nk} - \epsilon_{mk})^2-(\hbar\omega+\iota\eta)^2},
\end{aligned}
\end{equation}
where $n$, $m$ are band indexes, $\epsilon_n$ and $\epsilon_m$ are the eigenvalues, $f\textsubscript{nk}$ is the Fermi distribution function, BZ indicates the first Brillouin zone, $\hat{j}_x^{spinz}=\frac{1}{2}\{\hat{s_z},\hat{v_x}\}$ is the spin current operator and $\hat{s_z}=\frac{\hbar}{2}\sigma_{z}$ is the spin operator,
$\hat{v}_y = \frac{1}{\hbar}\frac{\partial H(k)}{\partial k_y}$ is the velocity operator and the frequency $\omega$ and $\eta$ are set to zero in the direct current (DC) clean limit.
The SHC for the Sb$_{0.5}$As$_{0.5}$ ordered alloy is shown in Fig.~\ref{fig5}. The value of SHC at Fermi energy is 330($\hbar$/e)(S/cm). One can see the SHC changes very sharply around Fermi energy, indicating that significant temperature dependence is expected. 
The SHC of the Sb$_{1-x}$As$_x$ ordered alloys range from 176-602 ($\hbar$/e)(S/cm). These values are lower than that predicted for Bi \cite{qu2023intrinsic}, yet comparable to those predicted for BiSb alloys, Bi$_2$Te$_3$, Bi$_2$Se$_3$, Sb$_2$Se$_3$, and Sb$_2$Te$_3$\cite{sharma2021light,farzaneh2020intrinsic}, as listed in Table II.

The SHC monotonically decreases with increased arsenic concentration. This is due to the weaker spin-orbit coupling strength of As compared to Sb and Bi. Still, we predict that strong spin-orbit coupling and Berry curvature in these materials produce large spin Hall conductivity, which can be used for the generation of fully spin-polarized current. The calculated spin Hall conductivities for all other concentrations in SbAs alloys are shown in Fig.~S3 in the Supplemental Material \cite{Zubair_supplement}. 

The in-plane lattice parameters of Sb$_{1-x}$As$_x$ ordered alloys display a linear behavior as function of composition $x$, and are compared to the (111) in-plane lattice parameters of conventional semiconductors as in Fig.~\ref{fig6}(a); values of $a$ and $c$ are listed in Table II in the Supplemental Material \cite{Zubair_supplement}.
We also computed the Fermi level of Sb$_{0.5}$As$_{0.5}$ with respect to the vacuum level and compare that with ionization potential and electron affinity of conventional semiconductors such as GaAs, AlAs, and Ge from Ref.~\onlinecite{hinuma2014band}. Due to the polar nature of the Sb$_{0.5}$As$_{0.5}$ along the $c$ direction, we used a slab of planar hexagonal structure with the same volume per formula unit as that of the ground state of the  Sb$_{0.5}$As$_{0.5}$ to determine the average electrostatic potential of the bulk with respect to the vacuum level.
The results are shown in Fig.~\ref{fig6}(b). We find that Sb$_{0.5}$As$_{0.5}$, with an in-plane lattice parameter only 0.6\% larger than that of GaAs, has a Fermi level within the band gap, at 0.2 eV below the conduction band of GaAs and about 0.4 eV below the conduction band of AlAs.
These results indicate the possibility of epitaxial growth of Sb$_{0.5}$As$_{0.5}$ on GaAs, AlGaAs, or AlAs, with small strains, and depending on the Al concentration in AlGaAs, the WPs in Sb$_{0.5}$As$_{0.5}$ can be placed near the conduction band or lower in the gap in Al-rich AlGaAs.  This will also allow  
sweeping through the Weyl points of Sb$_{0.5}$As$_{0.5}$ by gating through a heterojunction with III-V semiconductors, offering unique opportunities in designing quantum devices.

\section{Summary and conclusions}

Using first-principles calculations, we investigated the electronic structure of Sb$_{1-x}$As$_x$ ordered alloys, focusing on their non-trivial topology.  We find that an ordered Sb$_{0.5}$As$_{0.5}$ is energetically preferable, yet all ordered structures are slightly higher in energy and display Weyl semimetal phases, with 12 Weyl points. The Weyl points are very close to the Fermi level ($\sim$50 meV above it) so that Sb$_{0.5}$As$_{0.5}$ exhibits significant intrinsic spin-Hall conductivity. We also investigated the possible integration with conventional semiconductors, finding that Sb$_{0.5}$As$_{0.5}$ is almost lattice matched to GaAs and AlAs(111) and with the Fermi level and the Weyl points, within the band gap of the semiconductor, offering unique opportunities in hybrid quantum devices.

\vspace{5mm}
\section*{Acknowledgements}
\vspace{-1mm}
This work was supported by the National Science Foundation award \#OIA-2217786, and the use of Bridges-2 at PSC through allocation DMR150099 from the Advanced Cyberinfrastructure Coordination Ecosystem: Services \& Support (ACCESS) program, which is supported by the National Science Foundation grant nos.~2138259, 2138286, 2138307, 2137603, and 2138296, and the DARWIN computing system at the University of Delaware, which is supported by the NSF grant no.~1919839. S.K. acknowledges funding from the Laboratory Directed Research and Development Program (Project No. 800025) at Princeton Plasma Physics Laboratory under U.S. Department of Energy Prime Contract No. DE-AC02-09CH11466.

\bibliography{Bib_ReV}
\end{document}